\documentclass[a4paper]{article}

\usepackage{INTERSPEECH2022}

\title{Disentangled Latent Speech Representation for Automatic Pathological Intelligibility Assessment}

\name{Tobias Weise$^{1,2}$, Philipp Klumpp$^2$, Kubilay Can Demir$^1$, Andreas Maier$^2$, Elmar N\"oth$^2$, Björn Heismann$^2$, Maria Schuster$^3$, Seung Hee Yang$^1$}
\address{
	$^1$Speech {\small \&} Language Processing Lab. Friedrich-Alexander-Universit\"at Erlangen-N\"urnberg, Germany\\
	$^2$Pattern Recognition Lab. Friedrich-Alexander-Universit\"at Erlangen-N\"urnberg, Germany\\
	$^3$Department of Otorhinolaryngology, Head and Neck Surgery. Ludwig-Maximilians University, Munich, Germany
}
\email{tobias.weise@fau.de, seung.hee.yang@fau.de}

\begin{document}
	
	\maketitle
	
	\begin{abstract}
		Speech intelligibility assessment plays an important role in the therapy of patients suffering from pathological speech disorders. Automatic and objective measures are desirable to assist therapists in their traditionally subjective and labor-intensive assessments. In this work, we investigate a novel approach for obtaining such a measure using the divergence in disentangled latent speech representations of a parallel utterance pair, obtained from a healthy reference and a pathological speaker. Experiments on an English database of Cerebral Palsy patients, using all available utterances per speaker, show high and significant correlation values ($R=-0.9$) with subjective intelligibility measures, while having only minimal deviation ($\pm 0.01$) across four different reference speaker pairs. We also demonstrate the robustness of the proposed method ($R=-0.89$ deviating $\pm 0.02$ over 1000 iterations) by considering a significantly smaller amount of utterances per speaker. Our results are among the first to show that disentangled speech representations can be used for automatic pathological speech intelligibility assessment, resulting in a reference speaker pair invariant method, applicable in scenarios with only few utterances available.
	\end{abstract}
	\noindent\textbf{Index Terms}: auto-encoder, speech disentanglement, DTW, pathological speech intelligibility

	\begin{figure*}[t]
		\centering
		\includegraphics[width=\linewidth]{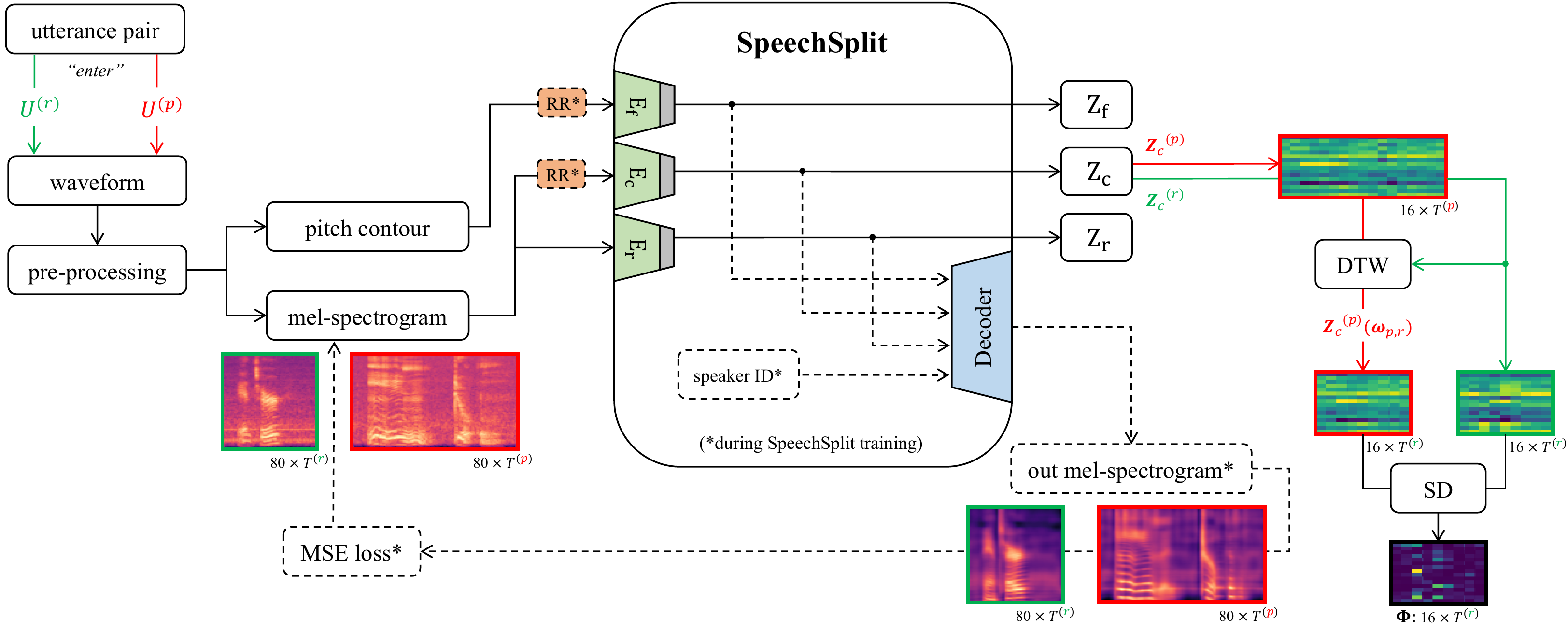}
		\caption{Schematic of proposed method, images are from a female healthy control and a UA-Speech patient with a subjective intelligibility score of 29\%. Considering $N$ utterances per speaker, matrices $\boldsymbol{\Phi}$ resulting from the SD operation are then used for the final intelligibility index calculation. The shown "out mel-spectrogram" is based on the three latent codes, disregarding timbre information.}
		\label{fig:schematic}
	\end{figure*}

	\section{Introduction}
	Speech disorders can impact different components required in the human speech production process, including phonation, articulation, and respiration \cite{natureSpeechProduction}. For example, dysarthria or dysphonia cause a patient's speech to become impaired, resulting in reduced intelligibility \cite{disorderDysarthria}. This causes communication difficulties in everyday life of the patients who suffer from incorrect articulation and voicing. Therefore, speech therapists typically use intelligibility assessment in order to characterize the severity of symptoms and to provide follow-up solutions \cite{dysaAssess, motorSpeechDisor}.\\
	As a result, perceptual evaluation performed in subjective listening tests is the most commonly used method for intelligibility assessment in clinical and therapeutic environments \cite{perceptualEval}. For an accurate assessment, human listeners are required to transcribe speech utterances, and intelligibility is evaluated based on the percentage of correctly understood words. However, this method is labor-intensive, expensive, and irreproducible. Also, this assessment is affected by the listener's familiarity with the patient's speech and linguistic context information in selected listening tests \cite{intell_listener_variability}. These drawbacks highlight the need for automatic and objective intelligibility measures to  assist clinicians in their assessments \cite{automatic_intell_important}.\\
	Objective metrics for intelligibility are also desirable in the speech enhancement domain, where they often deal with noisy and clean reference signals, such as the short-time objective intelligibility (STOI) measure \cite{STOI} and a temporal extension ESTOI \cite{ESTOI}. These measures require a reference signal to be present, which inspires a broad categorization \cite{spectralSub} of automatic speech intelligibility measures into two categories i) blind and ii) non-blind approaches, which can also be applied to the pathological setting. Here, the latter can utilize the fact that a reference signal is available, which enables exploiting information about intelligible speech from healthy speakers to obtain intelligibility measures. In contrast, blind approaches do not require reference signals and instead rely on automatically extracted features that are expected to correlate with intelligibility, which is followed by combining them via feature selection and regression training \cite{maier2009speech}. In the following paragraphs, we give examples for blind and non-blind approaches in the speech disorder domain, and propose our novel non-blind method.\\
	Blind approaches include the work of Vaysse et al. \cite{IntellHeadAndNeck} who aim at extracting parameters from the Envelope Modulation Spectrum that best account for the different levels of the speech signal's rhythmic structure, with subsequent feature selection and Support Vector Regression analysis to reveal the best parameter combination. Another example is the work of Falk et al. \cite{blind_1} who propose short- and long-term temporal dynamics measures and a composite measure, evaluated as correlates of subjective intelligibility. Moreover, methods that extract deep learning-based speaker embeddings like i-vectors or x-vectors have also been investigated for their usage in pathological intelligibility assessment \cite{iVector, iVector2, xVector}, where i-vectors represent the total variability subspace of a speaker and x-vectors aim to represent discriminative features between speakers.\\
	Non-blind approaches require healthy reference signals, for example to train an Automatic Speech Recognition (ASR) system to replace human listeners in the traditional subjective assessment. Pathological speech intelligibility is then computed based on word recognition rate. Such a system was successfully implemented by Maier et al. \cite{PEAKS} in an application run in hospitals to aid speech therapists in the assessment of patients. Another example of how healthy reference signals can be used is demonstrated by Janbakhshi et al. \cite{PSTOI_PESTOI}, called P-STOI and P-ESTOI. They adopted the already mentioned speech enhancement measures STOI and ESTOI in the pathological domain, based on creating an average healthy speaker reference and using time-alignment before computing their short-time objective intelligibility measure. The same authors proposed another approach \cite{spectralSub} where they use spectral subspace analysis by linear decomposition methods, i.e., Principal Component Analysis and Approximate Joint Diagonalization and computed the Grassman distance between a healthy and pathological representation as a content-independent objective intelligibility measure. Furthermore, Fritsch et al. \cite{julian} proposed an approach, which matches the respective posterior feature sequence of an utterance from a pathological speaker to a set of control speakers' utterances.\\
	In this paper, we investigate a novel approach of non-blind pathological speech intelligibility assessment based on the divergence in disentangled latent speech representations of a parallel utterance pair, obtained from a healthy reference and pathological speaker. Disentangled representations have already shown promising results in the field of speech pathology detection \cite{eFHVA}. Here, we adopt SpeechSplit \cite{SpeechSplit} to obtain such representations, which enables disregarding other (highly personal) information and we demonstrate that this approach delivers comparable results for a wide selection of healthy reference speakers, demonstrating a large degree of reference speaker invariance. Our implementation is available online\footnote{https://github.com/tobwei/disentIntel}.

	\section{SpeechSplit Architecture Overview}
	The authors of SpeechSplit hypothesize that speech information can be decomposed into four components: phonetic content $Z_c$, intonation/pitch $Z_f$, rhythm/timing $Z_r$, and lastly $timbre$, which is associated with the speaker identity. It was originally designed to achieve unsupervised (i.e., transcription-free) Voice Conversion (VC) using disentangled speech codes via a triple information bottleneck. The used architecture is an auto-encoder consisting of three encoders, extracting the above-mentioned three different latent codes and a decoder, reconstructing the original input mel-spectrogram based on these codes and a one-hot encoded speaker id, subject to a Mean Squared Error (MSE) loss function. The encoders receive different inputs, with the rhythm $E_r$ and the content $E_c$ encoders taking mel-spectrograms, and the frequency encoder $E_f$ taking the pitch contour as input. During training, the content and frequency encoder inputs are additionally subjected to a Random Resampling (RR) module, which intentionally disrupts the time domain. A mathematical proof is provided in the original paper on why this architecture design enforces speech disentanglement into the four components. An intuitive example for this is the RR module, which disrupts the time domain for $E_c$ and $E_f$, but not for $E_r$. Consequently, this makes $E_r$ the only encoder having access to the full time information and it will therefore focus on extracting a latent representation that encodes timing/rhythm information. Finally, the authors observed that SpeechSplit follows a "fill in the blank" mechanism, where the decoder fills blanks provided by the rhythm code with content and pitch information extracted by the relative encoders.

	\section{Pathological Intelligibility Assessment Based on Latent Speech Representation}
	From the three latent speech codes extracted by SpeechSplit, the content code $Z_c$ is the most promising for the task of pathological speech intelligibility assessment. Using this code is similar to the concept applied in the objective ASR approach when comparing a time-aligned pathological content code to a known reference. In contrast, $timbre$ information is speaker-dependent, which means that it does not contribute to pathology assessment and should therefore be disregarded. To validate the importance of $Z_c$, we also investigate the other two extracted codes ($Z_f$ and $Z_r$) using the same methodology. The following subsections describe code extraction, time alignment, reference speaker selection, and intelligibility index calculation.

	\subsection{Latent Speech Code Extraction}
	Since the proposed method is based on utilizing extracted encoder representations, our model has to be trained differently compared to the original SpeechSplit architecture, which was optimized for it's decoder output in the VC setting. This also manifests in the fact that model selection was originally exclusively done based on training loss, whereas we choose the final weights based on training and validation loss. SpeechSplit authors train their architecture with a small number of speakers and a relatively large amount of speech each. In contrast, and in order to achieve generalization properties of the encoders, we train with drastically more speakers and fewer speech data per speaker. Hyperparameters are the same as for the original architecture, except for padding parameters. Model input computations also follow the original SpeechSplit implementation, consisting of 80-dimensional mel-spectrograms with a frame length of 64 ms and 16 ms frame hop, as well as normalized and quantized pitch contours. Assuming a $80 \times T^{(x)}$ dimensional mel-spectrogram, computed from an utterance $U^{(x)}$, with $T^{(x)}$ being the total number of frames and $x \in \{r, p\}$ corresponding to a (healthy) reference and pathological speaker, then $\boldsymbol{Z}_c^{(x)}$ denotes a $C \times T^{(x)}$ dimensional content code. Here, $C$ is a hyperparameter of $E_c$, set to the original SpeechSplit bottleneck dimension of $16$. Next, a pair of such extracted latent codes from two different speakers are subjected to time-alignment.

	\subsection{Time Alignment}
	Based on a parallel utterance pair $ U^{(r)}, U^{(p)}$ from two different speakers, let $\boldsymbol{Z}_c^{(r)}, \boldsymbol{Z}_c^{(p)}$ denote the $ C \times T^{(r)}$ and $C \times T^{(p)}$ dimensional resulting content code pair, where $T^{(r)} \neq T^{(p)}$ in most cases because of individual speaking rates. Additionally, let $\boldsymbol{z}_c^{(r)}(t_i)$ and $\boldsymbol{z}_c^{(r)}(t_j)$ denote the content representation at time frame $t_i$ and $t_j$ respectively. These two codes are then time-aligned via the Dynamic Time Warping (DTW) algorithm \cite{DTW_1, DTW_2}. This technique results in a discrete match between existing elements of the two code sequences $\boldsymbol{Z}_c^{(r)}$ and $\boldsymbol{Z}_c^{(p)}$ by arranging them on a $T^{(r)} \times T^{(p)}$ grid, where each point $(i, j)$ corresponds to an alignment between elements $\boldsymbol{z}_c^{(r)}(t_i)$ and $\boldsymbol{z}_c^{(r)}(t_j)$. Minimizing the -- in this case Euclidean -- distance $\delta$ between these elements results in a mapping, which is denoted as warping path $\boldsymbol{\omega}= w_1, ..., w_k$. The problem of finding the warping path with the minimal distance measure can be expressed by $DTW(\boldsymbol{Z}_c^{(r)}, \boldsymbol{Z}_c^{(p)}) = min_{\boldsymbol{\omega}} \left[ \sum_{k=1}^{p}\delta(w_k) \right]$. As depicted by an example in Figure \ref{fig:schematic}, this ensures that the warped content codes $\boldsymbol{Z}_c^{(p)}(\boldsymbol{\omega}_{p,r})$ and $\boldsymbol{Z}_c^{(r)}(\boldsymbol{\omega}_{r,p})$ are element-wise aligned in the time domain.

	\subsection{Reference Speaker}
	The proposed method is partially inspired by the already mentioned STOI and ESTOI measures introduced in the speech enhancement domain. Here, the authors require time-aligned clean and noisy signals in order to predict intelligibility, which has already been adopted in the field of pathological speech intelligibility assessment as P-STOI/P-ESTOI, delivering state-of-the-art performance. However, the latter methods require the creation of an utterance-based average reference signal, created by randomly selecting one reference speaker to whom all other healthy speaker signals of the same utterance are first time-aligned, before taking the frame-wise average to create the final reference speech representation of a specific utterance. In contrast to this, the proposed method tries to alleviate this drawback by design and thus being reference speaker pair selection invariant. It is designed in a way that all information besides the phonetic content, extracted in the latent representation $Z_c$, is discarded before calculating the intelligibility index based on the difference of a reference and pathological signal. It should be noted that since speaker gender influences phonetic production \cite{phonetic_gender_1, phonetic_gender_2, phonetic_gender_3} we use a gender-matched reference speaker pair to their pathological counterparts to achieve best performance.

	\subsection{Intelligibility Index}
	Starting point for the following calculations, also depicted by an example in Figure \ref{fig:schematic}, are a pair of time aligned content codes $\boldsymbol{Z}_c^{(p)}(\boldsymbol{\omega}_{p,r})$, $\boldsymbol{Z}_c^{(r)}(\boldsymbol{\omega}_{r,p})$, extracted from a (healthy) reference and a pathological utterance. 
	\begin{equation}
		\label{eq:1}
		\boldsymbol{\Phi} = \left( \boldsymbol{Z}_c^{(r)}(\boldsymbol{\omega}_{r,p}) - \boldsymbol{Z}_c^{(p)}(\boldsymbol{\omega}_{p,r}) \right)^2 
	\end{equation}
	The result $\boldsymbol{\Phi}$ of Equation \eqref{eq:1} denotes a $C \times T$ dimensional and utterance $n \in N$ dependent representation of the squared differences (SD) in the corresponding aligned content codes. Considering a total of $N$ utterances, we define the intelligibility index
	\begin{equation}
		\label{eq:2}
		I = \frac{1}{CTN} \sum_{c=1}^{C} \sum_{t=1}^{T} \sum_{n=1}^{N} \phi_{c,t,n}
	\end{equation}
	of a (pathological) speaker as three dimensional average over a volume, consisting of $N$ stacked and utterance $n$ dependent difference matrices $\boldsymbol{\Phi}$, as seen in Equation \eqref{eq:2}.

	\section{Experimental Results}
	
	\subsection{Speech Corpora}
	
	\subsubsection{Training Corpora}
	SpeechSplit training was performed on a subset of the English Common Voice corpus \cite{CommonVoice}. Maintained by the Mozilla Corporation, this multilingual and publicly available voice dataset contains spoken text samples, contributed and evaluated by volunteers worldwide through the project's website. However, only text transcriptions are provided in terms of references, which poses a problem since SpeechSplit requires speaker id's for its training process. In order to circumvent this drawback, a subset of this corpus had to be created, following the process described by Klumpp et al. \cite{klumpp2022common} in order to connect speech files to a specific speaker, maintaining anonymity. The official website states that the English CV corpora contains almost 80,000 speakers, connecting files as described yields 14,212 identifiable speakers, with selecting 6-21 files (sampling at 16kHz) per speaker to solve the partially huge imbalance in speech contributions of different donors, resulting in 30-60 seconds of speech per speaker. Additionally, from all available files per speaker, we randomly reserve one file for the validation loss calculation, where 50/14,212 files are randomly selected per epoch. In contrast, SpeechSplit authors originally trained their model with 82 speakers from the VCTK \cite{VCTK} corpus, used approximately 15 minutes of speech per speaker, and performed model selection solely based on training loss.

	\subsubsection{Evaluation Corpora}
	Evaluation of the proposed method is performed on the UA-Speech corpus \cite{UASpeech}, comprised of 15 Cerebral Palsy patients (4 female, 11 male) and 13 healthy controls (4 female, 9 male). Each patient and control speaker read 765 isolated utterances, from which 760 are used for evaluation since 5 utterances were corrupted for some of the speakers. Subjective speech intelligibility scores of patients range from 2\% to 95\%. The pre-processing step depicted in Figure \ref{fig:schematic} required two steps for this corpora: first cutting 15\% at the beginning and end of the total length of an utterance recording and second, performing a subsequent Voice Activity Detection (VAD). The first step was necessary because there is often (especially for control speaker recordings) an audible mouse-click sound near the very end of a recording, which is picked up as a voiced segment by the VAD algorithm. Cutting in total 30\% of recording length is unproblematic for this corpus since recordings contain a lot of silence at the start and end. However, skipping the cutting step would be problematic for our proposed method, which is based on detecting content differences between two signals, where one contains a mouse-click and the other does not.

	\begin{figure}[]
		\centering
		\includegraphics[width=\linewidth]{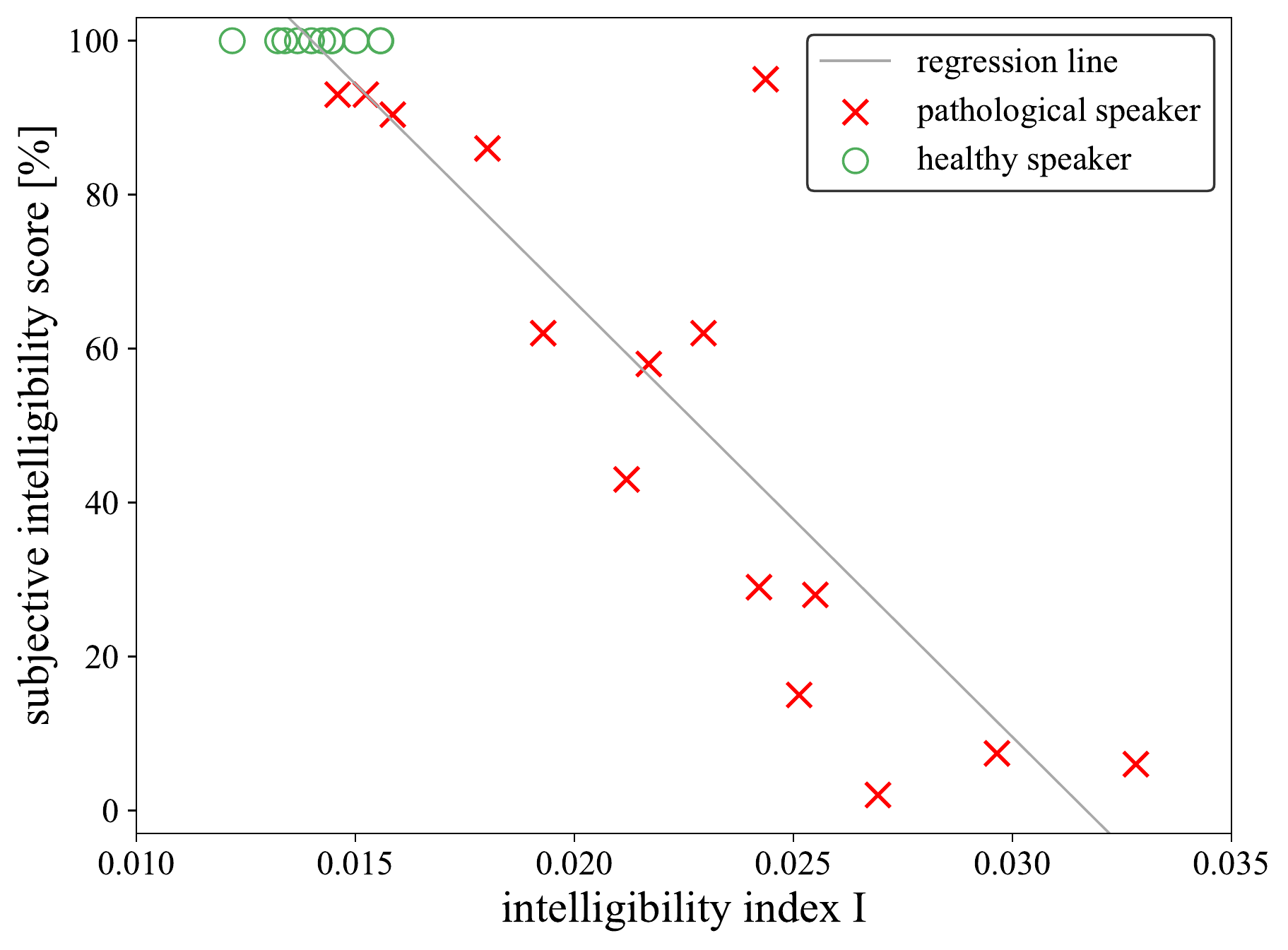}
		\caption{Scatter plot including least squares-regression line for a specific reference speaker pair, considering all 26 speakers with $R = -0.91, p = 6.8e\text{--}11$ and $R_s=-0.89, p=6e\text{--}10$.}
		\label{fig:scatter_plot}
	\end{figure}

	\subsection{Evaluation}
	As evaluation measures, Pearson correlation coefficient ($R$) and Spearman rank correlation coefficient ($R_s$) between the calculated intelligibility index $I$ and the subjective intelligibility scores, alongside their corresponding p-values (significance level), are computed. Additionally, a scatter plot is depicted to visualize this correlation, as well as a least-squares regression line. To calculate $\boldsymbol{\Phi}$ and finally the intelligibility index $I$ for each speaker in the UA-Speech corpus, mel-spectrogram computation follows the original SpeechSplit implementation, from which the corresponding content codes $\boldsymbol{Z}_c$ are then extracted by inferring with the best performing SpeechSplit model based on training and validation loss. In addition to all UA-Speech pathological speakers ($15$, "-pat"), the healthy controls except a gender-balanced pair selected as reference speakers ($(15 + (13 - 2) = 26)$, "-all") are also evaluated for their intelligibility index $I$, in which case both $U^{(r)}$ and $U^{(p)}$ contain healthy speech. In total 4 different reference speaker pairs were selected and evaluated against 15 pathological and 11 control speakers since there are only 4 female controls available.

	\begin{table}[]
		\caption{Correlation results, when considering pathological (15, "-pat") and also including healthy speakers (26, "-all"). Bold entries indicate significant correlations, i.e., $p < 0.05$.}
		\label{tab:results_numbers}
		\centering
		\begin{tabular}{lcccc}
			\toprule
			Measure & $R$ & $p$ & $R_S$ & $p$ \\
			\midrule 
			\multicolumn{4}{c}{\textbf{i) 20 utterances per speaker\textsuperscript{$*)$}}} \\
			\midrule
			Z\textsubscript{c}-all& $\boldsymbol{-0.89}$ & $1.1e\text{--}6$ & $\boldsymbol{-0.86}$ & $1.5e\text{--}5$ \\
			& $\pm 0.02$& & $\pm 0.03$ & \\
			Z\textsubscript{c}-pat& $\boldsymbol{-0.81}$ & $0.04$ & $\boldsymbol{-0.76}$ & $0.04$ \\
			& $\pm 0.05$& & $\pm 0.07$ & \\
			\midrule
			\multicolumn{4}{c}{\textbf{ii) 760 utterances per speaker\textsuperscript{$**)$}}} \\
			\midrule
			Z\textsubscript{f}-all& $-0.27$ & $0.18$ & $-0.13$ & $0.52$ \\
			Z\textsubscript{r}-all& $\boldsymbol{-0.64}$ & $4.6e\text{--}4$ & $\boldsymbol{-0.56}$ & $3e\text{--}3$ \\
			\midrule
			Z\textsubscript{c}-all& \boldsymbol{$-0.90$} & $6.6e\text{--}10$ & $\boldsymbol{-0.88}$ & $6.7e\text{--}9$ \\
			& $\pm 0.01$& & $\pm 0.01$ &  \\
			Z\textsubscript{c}-pat& \boldsymbol{$-0.83$} & $1.8e\text{--}4$ & \boldsymbol{$-0.80$} & $1.2e\text{--}3$ \\
			& $\pm 0.01$ & & $\pm 0.03$ & \\
			\bottomrule
		\end{tabular}
		\vspace{0.1ex}
		{\\$*)$ \footnotesize provided deviations relate to 1000 iterations} \\
		{$**)$ \footnotesize provided deviations relate to different reference speaker pairs}
	\end{table}

	\subsection{Results}
	The least-squares regression line in Figure \ref{fig:scatter_plot} is computed considering all speakers. However, it looks very similar when considering only pathological speakers. It can be observed that there is a visible separation between the pathological and healthy speakers, with the latter forming a cluster. A short investigation has been conducted to investigate the visible outlier in this scatter plot, with a possible explanation of some distinct background sounds in the recordings of this female patient, which could be picked up by the VAD algorithm as voiced segments and thus explain the resulting high squared differences and thus high intelligibility index $I$.\\
	Results shown in Table \ref{tab:results_numbers} are the correlation values based on few i) $N = 20$ and maximally available ii) $N = 760$ utterances per speaker for the intelligibility index $I$ calculation. Presented correlation results in these two sections are mean and standard deviation values for 1000 iterations and four different reference speaker pair experiments respectively, with p-values being the worst performing ones obtained in terms of significance level. Scenario ii) shows that the proposed method achieves high and significant Pearson and Spearman correlation coefficients, while also achieving reference speaker invariance. However, results also indicate that scenario i) can already yield good and significant results, by considering significantly fewer utterances. Overall, results shown in Table \ref{tab:results_numbers} indicate that it is beneficial to include healthy speakers.\\
	Lastly, $Z_f$ and $Z_r$ code results are also presented in Table \ref{tab:results_numbers}, where the latter correlation value can be explained by the aforementioned "fill in the blank" mechanism, with the rhythm code providing blanks to be filled with content and frequency code.

	\section{Conclusions}
	We have investigated an automatic and non-blind (reference signal is available) approach towards objective pathological intelligibility assessment, based on the divergence in disentangled latent speech representations of a parallel utterance pair, obtained from a healthy reference and a pathological speaker. It was shown that the proposed method is reference speaker pair selection invariant and can deliver results with only few available utterances per speaker, which is appealing from a practical point of view. Possible future directions include training a model with multilingual data, enabling evaluation on more pathological speech databases. 
	
	\section{Acknowledgements}
	We gratefully acknowledge funding for this study by Friedrich-Alexander-University Erlangen-Nuermberg, Medical Valley e.V. and Siemens Healthineers AG within the framework of d.hip campus. 
	
	\bibliographystyle{IEEEtran}
	
	\bibliography{mybib}

\begin{thebibliography}{10}
\providecommand{\url}[1]{#1}
\csname url@samestyle\endcsname
\providecommand{\newblock}{\relax}
\providecommand{\bibinfo}[2]{#2}
\providecommand{\BIBentrySTDinterwordspacing}{\spaceskip=0pt\relax}
\providecommand{\BIBentryALTinterwordstretchfactor}{4}
\providecommand{\BIBentryALTinterwordspacing}{\spaceskip=\fontdimen2\font plus
\BIBentryALTinterwordstretchfactor\fontdimen3\font minus
  \fontdimen4\font\relax}
\providecommand{\BIBforeignlanguage}[2]{{%
\expandafter\ifx\csname l@#1\endcsname\relax
\typeout{** WARNING: IEEEtran.bst: No hyphenation pattern has been}%
\typeout{** loaded for the language `#1'. Using the pattern for}%
\typeout{** the default language instead.}%
\else
\language=\csname l@#1\endcsname
\fi
#2}}
\providecommand{\BIBdecl}{\relax}
\BIBdecl

\bibitem{natureSpeechProduction}
G.~Hickok, ``Computational neuroanatomy of speech production,'' \emph{Nature
  reviews neuroscience}, vol.~13, no.~2, pp. 135--145, 2012.

\bibitem{disorderDysarthria}
P.~Enderby, ``Disorders of communication: dysarthria,'' \emph{Handbook of
  clinical neurology}, vol. 110, pp. 273--281, 2013.

\bibitem{dysaAssess}
------, ``Frenchay dysarthria assessment,'' \emph{British Journal of Disorders
  of Communication}, vol.~15, no.~3, pp. 165--173, 1980.

\bibitem{motorSpeechDisor}
A.~Lowit and R.~D. Kent, \emph{Assessment of motor speech disorders}.\hskip 1em
  plus 0.5em minus 0.4em\relax Plural publishing, 2010.

\bibitem{perceptualEval}
S.~Fex, ``Perceptual evaluation,'' \emph{Journal of voice}, vol.~6, no.~2, pp.
  155--158, 1992.

\bibitem{intell_listener_variability}
M.~McHenry, ``An exploration of listener variability in intelligibility
  judgments,'' 2011.

\bibitem{automatic_intell_important}
C.~Middag, ``Automatic analysis of pathological speech,'' Ph.D. dissertation,
  Ghent University, 2012.

\bibitem{STOI}
C.~H. Taal, R.~C. Hendriks, R.~Heusdens, and J.~Jensen, ``A short-time
  objective intelligibility measure for time-frequency weighted noisy speech,''
  in \emph{2010 IEEE International Conference on Acoustics, Speech and Signal
  Processing}, 2010, pp. 4214--4217.

\bibitem{ESTOI}
J.~Jensen and C.~H. Taal, ``An algorithm for predicting the intelligibility of
  speech masked by modulated noise maskers,'' \emph{IEEE/ACM Transactions on
  Audio, Speech, and Language Processing}, vol.~24, no.~11, pp. 2009--2022,
  2016.

\bibitem{spectralSub}
P.~Janbakhshi, I.~Kodrasi, and H.~Bourlard, ``Spectral subspace analysis for
  automatic assessment of pathological speech intelligibility.'' in
  \emph{INTERSPEECH}, 2019, pp. 3038--3042.

\bibitem{maier2009speech}
A.~Maier, \emph{Speech of children with cleft lip and palate: Automatic
  assessment}.\hskip 1em plus 0.5em minus 0.4em\relax Logos-Verlag, 2009.

\bibitem{IntellHeadAndNeck}
R.~Vaysse, J.~Farinas, C.~Astésano, and R.~Andre-Obrecht, ``Automatic
  extraction of speech rhythm descriptors for speech intelligibility assessment
  in the context of head and neck cancers,'' 08 2021, pp. 1912--1916.

\bibitem{blind_1}
T.~H. Falk, R.~Hummel, and W.-Y. Chan, ``Quantifying perturbations in temporal
  dynamics for automated assessment of spastic dysarthric speech
  intelligibility,'' in \emph{2011 IEEE International Conference on Acoustics,
  Speech and Signal Processing (ICASSP)}, 2011, pp. 4480--4483.

\bibitem{iVector}
D.~Mart{\'\i}nez, E.~Lleida, P.~Green, H.~Christensen, A.~Ortega, and
  A.~Miguel, ``Intelligibility assessment and speech recognizer word accuracy
  rate prediction for dysarthric speakers in a factor analysis subspace,''
  \emph{ACM Transactions on Accessible Computing (TACCESS)}, vol.~6, no.~3, pp.
  1--21, 2015.

\bibitem{iVector2}
I.~Laaridh, C.~Fredouille, A.~Ghio, M.~Lalain, and V.~Woisard, ``Automatic
  evaluation of speech intelligibility based on i-vectors in the context of
  head and neck cancers,'' in \emph{Interspeech}.\hskip 1em plus 0.5em minus
  0.4em\relax ISCA, 2018, pp. 2943--2947.

\bibitem{xVector}
S.~Quintas, J.~Mauclair, V.~Woisard, and J.~Pinquier, ``Automatic prediction of
  speech intelligibility based on x-vectors in the context of head and neck
  cancer.'' in \emph{INTERSPEECH}, 2020, pp. 4976--4980.

\bibitem{PEAKS}
A.~Maier, T.~Haderlein, U.~Eysholdt, F.~Rosanowski, A.~Batliner, M.~Schuster,
  and E.~Nöth, ``Peaks – a system for the automatic evaluation of voice and
  speech disorders,'' \emph{Speech Communication}, vol.~51, no.~5, pp.
  425--437, 2009.

\bibitem{PSTOI_PESTOI}
P.~Janbakhshi, I.~Kodrasi, and H.~Bourlard, ``Pathological speech
  intelligibility assessment based on the short-time objective intelligibility
  measure,'' in \emph{ICASSP 2019 - 2019 IEEE International Conference on
  Acoustics, Speech and Signal Processing (ICASSP)}, 2019, pp. 6405--6409.

\bibitem{julian}
J.~Fritsch and M.~Magimai-Doss, ``Utterance verification-based dysarthric
  speech intelligibility assessment using phonetic posterior features,'' 01
  2021.

\bibitem{eFHVA}
J.~Qi \emph{et~al.}, ``Speech disorder classification using extended factorized
  hierarchical variational auto-encoders,'' \emph{arXiv preprint
  arXiv:2106.07337}, 2021.

\bibitem{SpeechSplit}
K.~Qian, Y.~Zhang, S.~Chang, M.~Hasegawa-Johnson, and D.~Cox, ``Unsupervised
  speech decomposition via triple information bottleneck,'' in
  \emph{Proceedings of the 37th International Conference on Machine Learning},
  ser. Proceedings of Machine Learning Research, H.~D. III and A.~Singh, Eds.,
  vol. 119.\hskip 1em plus 0.5em minus 0.4em\relax PMLR, 13--18 Jul 2020, pp.
  7836--7846.

\bibitem{DTW_1}
D.~J. Berndt and J.~Clifford, ``Using dynamic time warping to find patterns in
  time series.'' in \emph{KDD workshop}, vol.~10, no.~16.\hskip 1em plus 0.5em
  minus 0.4em\relax Seattle, WA, USA:, 1994, pp. 359--370.

\bibitem{DTW_2}
\BIBentryALTinterwordspacing
\emph{Dynamic Time Warping}.\hskip 1em plus 0.5em minus 0.4em\relax Berlin,
  Heidelberg: Springer Berlin Heidelberg, 2007, pp. 69--84. [Online].
  Available: \url{https://doi.org/10.1007/978-3-540-74048-3\_4}
\BIBentrySTDinterwordspacing

\bibitem{phonetic_gender_1}
J.~Pardo, A.~Urmanche, S.~Wilman, and J.~Wiener, ``Phonetic convergence and
  talker sex: It’s complicated,'' \emph{The Journal of the Acoustical Society
  of America}, vol. 139, no.~4, pp. 2105--2106, 2016.

\bibitem{phonetic_gender_2}
Y.~Samuelsson, ``Gender effects on phonetic variation and speaking styles,''
  2007.

\bibitem{phonetic_gender_3}
E.~Oh, ``Effects of speaker gender on voice onset time in korean stops,''
  \emph{Journal of Phonetics}, vol.~39, no.~1, pp. 59--67, 2011.

\bibitem{CommonVoice}
R.~Ardila, M.~Branson, K.~Davis, M.~Henretty, M.~Kohler, J.~Meyer, R.~Morais,
  L.~Saunders, F.~M. Tyers, and G.~Weber, ``Common voice: {A}
  massively-multilingual speech corpus,'' \emph{CoRR}, vol. abs/1912.06670,
  2019.

\bibitem{klumpp2022common}
P.~Klumpp, T.~Arias-Vergara, P.~A. P{\'e}rez-Toro, E.~N{\"o}th, and J.~R.
  Orozco-Arroyave, ``Common phone: A multilingual dataset for robust acoustic
  modelling,'' \emph{arXiv preprint arXiv:2201.05912}, 2022.

\bibitem{VCTK}
C.~Veaux, Y.~Junichi, and K.~MacDonald, ``Superseded-cstrv vctk corpus: English
  multi-speaker corpus for cstr voice cloning toolkit,'' \emph{University of
  Edinburgh. The Centre for Speech Technology Research (CSTR)}, 2016.

\bibitem{UASpeech}
H.~Kim, M.~Hasegawa-Johnson, A.~Perlman, J.~Gunderson, K.~Watkin, and S.~Frame,
  ``Dysarthric speech database for universal access research,'' 01 2008, pp.
  1741--1744.

\end{thebibliography}
\end{document}